\documentclass[graybox]{svmult}
\usepackage{graphicx} % Required for inserting images
\usepackage{amsmath, amsfonts, amsthm}
\usepackage{lmodern}
\usepackage[T1]{fontenc}
\usepackage{mathrsfs}
\usepackage{xcolor}
\usepackage{hyperref, caption}

\def \arcsec      {\text{$^{\prime\prime}$}}
\newcommand{\aap}{A\&A} % Astronomy & Astrophysics
\newcommand{\mnras}{MNRAS} % Monthly Notices of the Royal Astronomical Society
\newcommand{\nat}{Nature}
\newcommand{\araa}{Annual Review of Astronomy and Astrophysics}
\newcommand{\nar}{New Astronomy Reviews}
\newcommand{\apjs}{The Astrophysical Journal Supplement Series}
\newcommand{\pasa}{Publications of the Astronomical Society of Australia}
\begin{document}
\title*{Ten years of searching for relics of AGN jet feedback through RAD@home citizen science}
\author{Ananda Hota$^{1,2}$, Pratik Dabhade$^{3,4,5,2}$, Prasun Machado$^{2}$, Avinash Kumar$^{2}$, Ck. Avinash$^{2}$, Ninisha Manaswini$^{2}$, Joydeep Das$^{2}$, Sagar Sethi$^{2}$, Sumanta Sahoo$^{2}$, Shilpa Dubal$^{2}$, Sai Arun Dharmik Bhoga$^{2}$, P. K. Navaneeth$^{2}$, C. Konar$^{6,2}$, Sabyasachi Pal$^{7,2}$, Sravani Vaddi$^{2}$, Prakash Apoorva$^{2}$, Megha Rajoria$^{2}$, and Arundhati Purohit$^{2}$}
\institute{(1) UM-DAE Centre for Excellence in Basic Sciences, University of Mumbai, Mumbai-400098, India \\
(2) RAD@home Astronomy Collaboratory, Kharghar, Navi Mumbai-410210, India \\
(3) Instituto de Astrof\' isica de Canarias, Calle V\' ia L\'actea, s/n, E-38205, La Laguna, Tenerife, Spain \\ 
(4) Universidad de La Laguna (ULL), Departamento de Astrofisica, La Laguna, E-38206, Tenerife, Spain \\
(5) Astrophysics Division, National Centre for Nuclear Research, Pasteura 7, 02-093 Warsaw, Poland\\
(6) Amity Institute of Applied Sciences, Amity University Uttar Pradesh, Sector-125, Noida-201303, India \\
(7) Midnapore City College, Kuturia, Bhadutala, Paschim Medinipur, West Bengal-721129, India}
\maketitle
\abstract{Understanding the evolution of galaxies cannot exclude the important role played by the central supermassive black hole and the circumgalactic medium (CGM). Simulations have strongly suggested the negative feedback of AGN Jet/wind/outflows on the ISM/CGM of a galaxy leading to the eventual decline of star formation. However, no ``smoking gun'' evidence exists so far where relics of feedback, observed in any band, are consistent with the time scale of a major decline in star formation, in any sample of galaxies. Relics of any AGN-driven outflows will be observed as a faint and fuzzy structure which may be difficult to characterise by automated algorithms but trained citizen scientists can possibly perform better through their intuitive vision with additional heterogeneous data available anywhere on the Internet. RAD@home, launched on 15$^{\rm th}$ April 2013, is not only the first Indian citizen Science Research (CSR) platform in astronomy but also the only CSR publishing discoveries using any Indian telescope. We briefly report 11 CSR discoveries collected over the last eleven years. While searching for such relics we have spotted cases of offset relic lobes from elliptical and spiral, episodic radio galaxies with overlapping lobes as the host galaxy is in motion, large diffuse spiral-shaped emission, cases of jet-galaxy interaction, kinks and burls on the jets, a collimated synchrotron thread etc. Such exotic sources push the boundaries of our understanding of classical Seyferts and radio galaxies with jets and the process of discovery prepares the next generation for science with the upgraded GMRT and Square Kilometre Array Observatory(SKAO). }
\section{Introduction}
The current understanding of hierarchical structure formation, galaxy merger and feedback from the active galactic nuclei (AGN) has been widely accepted as prime drivers of galaxy evolution \cite{DiMatteo2005,Hopkins2008}. Incorporation of this AGN-feedback could explain various statistical correlations observed in large populations of galaxies like M-$\sigma$ relation, colour-magnitude relation, luminosity function etc. \cite{2012McNamara,2012Fabian,Kormendy2013}. During the galaxy merger, the nuclear star formation rate increases due to the infall of gas leading to high galactic superwind outflow as well as radiation, and radio-jet outflows from the central AGN \cite{2005Veilleux,2006Croton,Hardcastle2020}. These lead to the loss of cold gas and dust content in the merger remnant and eventually transform blue-star-forming late-type galaxies quickly to a red-looking early-type galaxy population. This quick loss within the merger timescale of nearly a billion years shows up as the gap between the ``blue cloud'' and ``red sequence'' of galaxies in the colour-magnitude diagram \cite{Heckman2014}

\begin{figure}
\centering
\includegraphics[width=0.525\columnwidth]{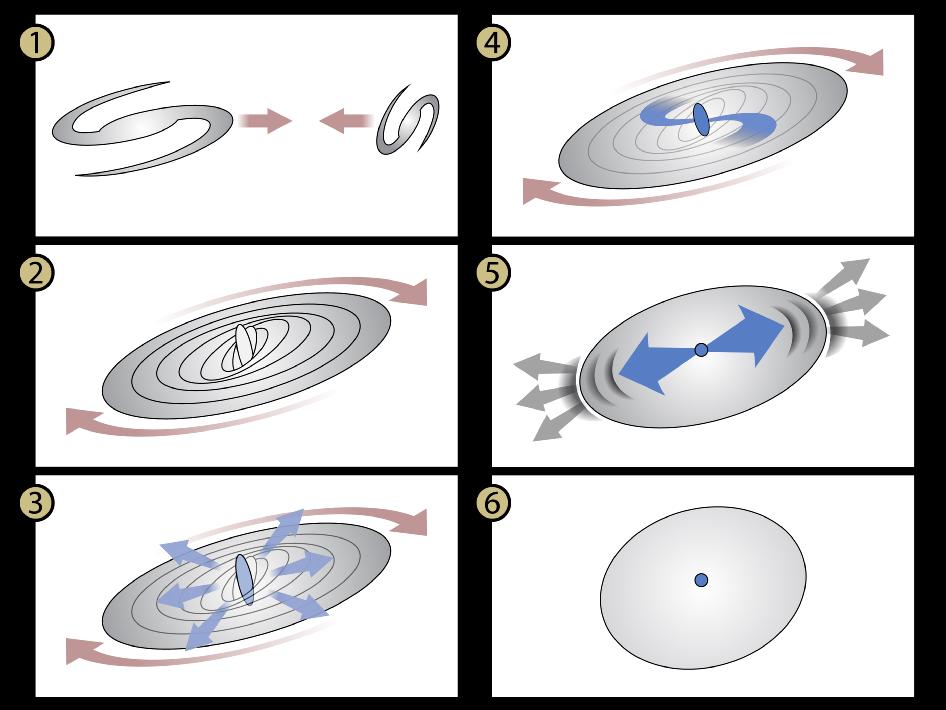}
\includegraphics[width=0.4\columnwidth]{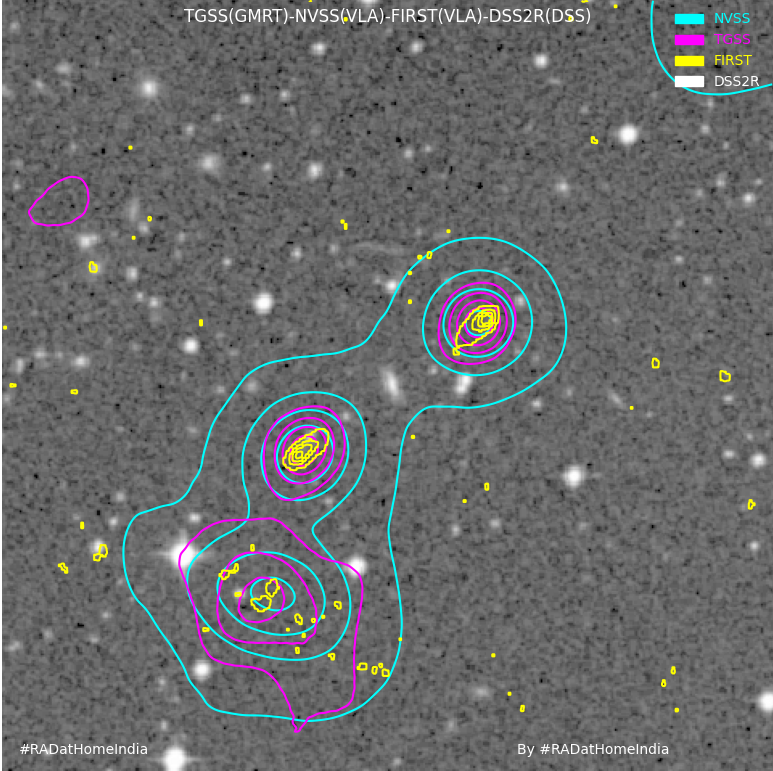}
\caption{ \small{ Left panel (NGC3801): Evolutionary scenario of NGC3801 adapted from \cite{Hota2012} and associated Press Release  \url{https://www.jpl.nasa.gov/news/cosmic-leaf-blower-robs-galaxy-of-star-making-fuel}. Right Panel (Speca): A typical RAD@home RGB-maker Composite Contour image where TGSS (magenta), NVSS(cyan) and FIRST(yellow) are plotted over DSS2R optical image. Contour levels (Jy beam$^{-1}$) are TGSS:[0.015, 0.031 0.046, 0.062] FIRST:[0.0005, 0.0026, 0.0048, 0.0069] NVSS:[0.0015, 0.01, 0.0185, 0.0271].
}}
\label{fig:a}
\end{figure}

However, this widely accepted standard model of AGN-feedback or outflows causing the decline of star formation as a process has not been observed which can be called ``smoking gun'' evidence. Massive outflows have been observed, power in it has been estimated to be sufficient, post-merger and post-starburst galaxies have been known, but in no sample so far the relics of the outflow match with the time-scale of decline of star formation \cite{Hota2016}. For example, in the post-starburst galaxies, an abrupt decline has been observed in the last one billion years. Massive molecular gas outflows have been observed in various types of galaxies like mergers, post-merger, post-starburst etc. In many cases these outflow rates were found to be significantly above the current star formation rates demonstrating the power of the feedback to cause the eventual decline of star formation following galaxy mergers. However, these outflow time scales are only a few tens of million years \cite{2005Veilleux}. Thus, due to the difference in time scales involved, the truncation of star formation (since a billion years) could not have been caused by the observed outflows (since a few tens of million years of outflows). Hence, there is a need to observe relics of AGN-driven outflow that could be older than the time scale since the decline of star formation in some transitional galaxies. Unlike in any other bands, evidence of multiple episodes of AGN activity is quite common in radio bands, seen clearly as `Double Double Radio Galaxies' (DDRG; see review by \cite{SaikiaDDRG}) with well-separated synchrotron age of the episodes of jets \cite{Konar2006}. In optical bands, the optical relics of quasar ionisation, popularly known as Hanny’s Voorwerps, have been found to be a few to ten thousand years old. In the first such objects with low radio frequency follow-up, it has been found that the radio outflow is 100 Million years old that is much older than the optical relic (10 thousand years) and the quasar is no longer active \cite{Smith2022}. This demonstrates the large scope in radio bands for the discovery of relic radio lobes and investigates the possible break in the star formation history of the AGN-host to support AGN-feedback models.
As an example, we briefly explain here the case of the Cosmic Leaf Blower galaxy, a post-merger galaxy with sub-galactic scale s-shaped radio lobes, NGC3801 \cite{Hota2012} with a speculative evolutionary scenario (Figure.~\ref{fig:a}, Left Panel). Over a billion years back an unequal mass merger between two spiral galaxies (Stage-1) led to the formation of a kinematically decoupled core (KDC) galaxy. The outer gaseous/stellar (H\,{\sc i}  and H$\alpha$) disk (r=15 kpc) rotates (v=280 km s$^{-1}$) along the major axis roughly aligned east-west (Stage-2). This outer gas disk shows some residual young star formation in UV, having an age of nearly 100-500 Million years while most of the galaxy looks yellowish, suggesting a typical early-type galaxy. The molecular gas (and dust) in the inner 4 kpc region rotate orthogonal to the larger scale gas disk. The radio jet emanates orthogonal to the central dust disk but after the hotspot regions, gradually bends to form an S-shape. The synchrotron age of these radio jets suggests it to be very young around 2.4 million years (Stage-4). Surrounding the hotspot regions, nearly 2 kpc away from the nucleus, X-ray emitting shock shells have been found which expand at nearly 850 km s$^{-1}$ speed (Stage-5). Since the jet is expanding in the plane of the sky the shock shell would reach the outer gas disk with some young star formation in the next 10 million years which is like a blink compared to the 330 Million year time-scale of rotation of the large-scale gas disk. So in the next ten million years into the future, the outer young-star-forming gas disk may be disrupted but the overall red colour of the galaxy can not be due to the ongoing AGN-driven radio/ionised-gas outflow/feedback and must be from an earlier episode (Stage-3). Thus, if deep low-frequency observation can find a relic radio emission which is around 100-500 million years old, only then, it can be a convincing case of ``smoking gun'' evidence (see \cite{Hota2012} for details).

\section{Discovery methods and Results}

Since the launch of RAD@home citizen science research collaboratory on 15th April 2013, citizen scientists have been trained to identify host galaxies and active/relic lobes of radio galaxies by making radio-optical overlays \cite{Hota2016}. First, they were using NASA Skyview (\url{https://skyview.gsfc.nasa.gov/}) for such overlays and later a Python-based RGB-maker web tool (\url{https://www.radathomeindia.org/rgbmaker}), launched by RAD@home on 26th January 2021 \cite{Kumar2023}. RAD@home training programmes [known as RAD@home Discovery Camps (week-long) and RAD@home Astronomy Workshops (one-day event)] were conducted in collaboration with research and educational institutes. After learning to characterise standard radio galaxies (FR I, FR II, wide angle tailed. head-tailed, DDRG, X-shaped radio galaxy etc.) they were asked to identify radio sources from Giant Metrewave Radio Telescope (GMRT) 150 MHz data (TGSS DR5 \& TGSS ADR1; \cite{Intema2017}) which does not fit into standard classes and/or are ``faint and fuzzy'' which could be potential cases of relic lobes of past AGN feedbacks. An example of how a comparison of bright NVSS radio emission blobs would be seen faintly in the FIRST data for the relic radio lobe of the episodic radio galaxy Speca \cite{Hota2011} is shown in Figure.~\ref{fig:a} (right panel). In comparison to NVSS the TGSS data being nearly ten times longer in wavelength would also be ten times brighter for a steep spectral relic radio lobe, assuming a spectral index of -1. Thus such a relic lobe detected as an NVSS blob seen at 3$\sigma$ (1.5 mJy) has a good chance to be seen in the TGSS at 3$\sigma$ at 15 mJy. Such thumb rules were helpful for citizen scientists without much understanding of angular scale-sensitive radio interferometric images. Such scientifically interesting sources were then followed up with the GMRT telescope through a multi-cycle proposal titled ``GMRT Observation of Objects Discovered by RAD@home Astronomy Collaboratory (GOOD-RAC)''. One of the potential demonstrators was the discovery of RAD-12, a unique AGN with a one-sided radio jet hitting a companion galaxy and bouncing back to form a radio bubble \cite{Hota2022}. Interestingly, unlike Minkowski's object, the companion galaxy shows no sign of enhanced star formation (positive feedback). Here we briefly report some exotic cases where relic radio lobe emissions have been identified which are currently being studied by the collaboratory [Hota et al. 2024 (in preparation)].

%\section{Notes on individual sources}
%\subsection{RAD-Q-mark galaxy}
\begin{figure}
\centering
\includegraphics[width=\columnwidth]{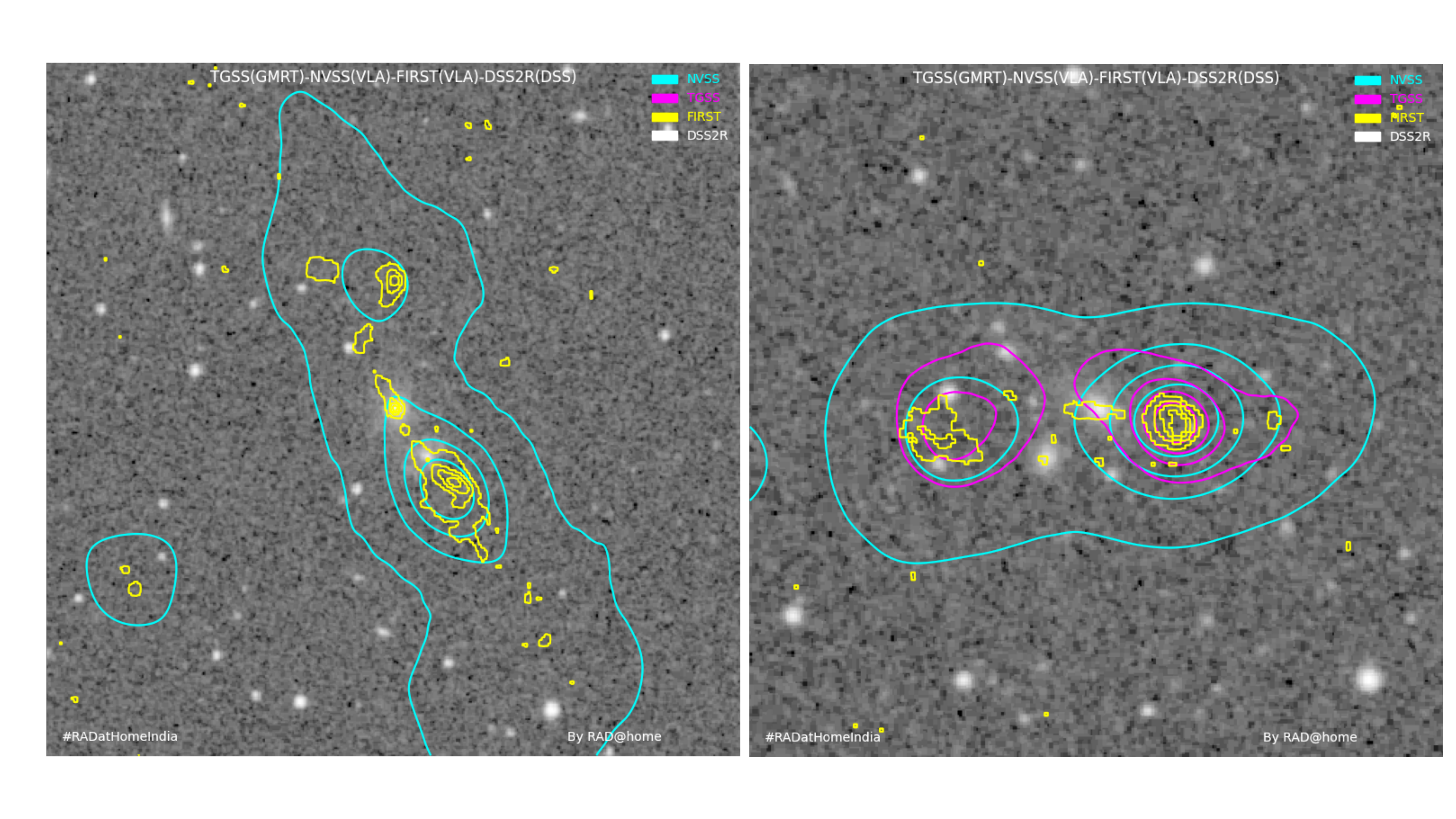}
\caption{\small{Left Panel (RAD-Q-mark galaxy): Contour levels (Jy beam$^{-1}$) are FIRST:[0.0005, 0.0023, 0.004, 0.0058], NVSS:[0.0015, 0.0262, 0.051, 0.0757], TGSS (not available). Right Panel (RAD-jet-burl): Contour levels (Jy beam$^{-1}$) are TGSS:[0.015, 0.04, 0.064, 0.089], FIRST: [0.0005, 0.0021, 0.0037, 0.0054], NVSS: [0.0015, 0.0119, 0.0223, 0.0328].}}
\label{fig:b}
\end{figure}

{\bf RAD-Q-mark galaxy:} The first practical skill acquired by participants of the RAD@home Discovery Camp, hosted at various institutes, is the ability to identify the host galaxy of a radio source.
% The first practical skill that participants of RAD@home Discovery Camp, hosted at different institutes, gain is identifying the host galaxy of a radio source.
For a FR I radio galaxy the radio peak matches with an optical peak but for FR II the radio peaks will be on either side of an optical galaxy. RAD12 was a peculiar case where the radio peak was in between two optical galaxies. RAD12, on follow-up observations with the GMRT, turned out to be a case where the one-sided jet hits the companion galaxy and returns to form a bubble (see  \cite{Hota2022} for details). Similarly, a radio source was found by a trained citizen scientist where the NVSS radio peak coincides roughly with a spiral galaxy, which was suspected as a Speca – spiral-host large episodic radio galaxy \cite{Hota2011}. However, the high-resolution FIRST data identifies the core jet with the neighbouring elliptical galaxy (RA: 09:18:59.38 Dec: +31:51:40.78), confirming that it is the host (Figure.~\ref{fig:b} left panel). Interestingly, the FIRST image looks like a question mark sign ``?''. So this Q-mark galaxy clearly shows compact emission, typically seen in younger radio jets/ lobes but also shows fainter diffuse emission in NVSS, extending beyond the bright/compact features. While the structure seen in FIRST on the northern side suggests a `kink' \cite{2022Dabhade} the brightened blob on the southern jet, coinciding spiral galaxy, suggests a possible jet-galaxy interaction, without any optical signature, as seen in RAD12. Recent LOFAR-Two-Metre-Sky-Survey data release 2  (LoTSS DR-2 data \cite{lotssdr2}) of the Q-mark galaxy confirms the structures with higher resolution and higher significance level, strongly supporting our interpretations. A detailed study of the target is currently being drafted.

{\bf RAD-jet-burl:} An unusual burl-like growth on the radio jets, other than the lobe and hotspot, was found during a RAD@home e-class conducted online during weekends (Date: 2-7 Oct. 2023). This is unlike the C- or U-shaped kink seen in the previous cases \cite{2022Dabhade}. For the burl seen on the western side of the host galaxy, there is no apparent optical galaxy in the location to suggest possible jet-galaxy interaction and the corresponding structure on the eastern lobe is also missing (Figure.~\ref{fig:b} right panel). This burl region is even brighter than the entire lobe on the other side. Given the redshift ($z=$0.140) of the host galaxy (RA: 14 33 40.754 Dec.: +63 58 30.65), clearly showing twin FRI jets in the FIRST data, the burl size can be as big as 25$\arcsec$ (60 kpc). Recent LoTSS data clearly shows the burl seen here but it lacks any internal C-/U-shaped structure. A resolved spectral index and polarisation image of the jet-burl would be helpful in understanding its nature and origin.

%\subsection{RAD-Rabbit}
\begin{figure}
\centering
\includegraphics[width=\columnwidth]{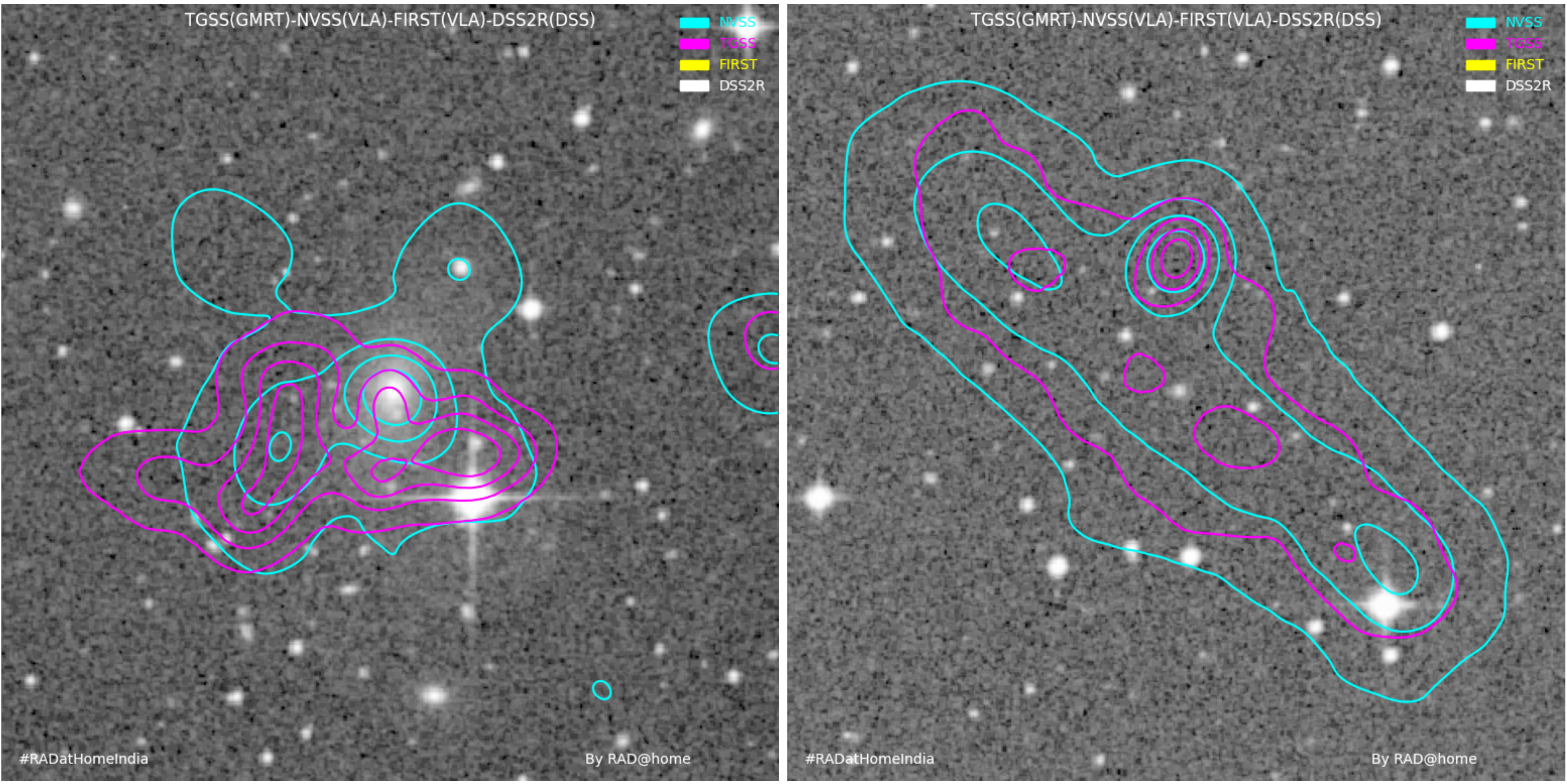}
\caption{\small{Left panel (RAD-Rabbit): Contour levels (Jy beam$^{-1}$) are TGSS:[0.015, 0.038, 0.061, 0.083] NVSS: [0.0015, 0.0066, 0.0117, 0.0169], FIRST (not available). Right Panel (RAD-giant DDRG): Contour levels (Jy beam$^{-1}$) are TGSS:[0.015 0.058 0.1 0.143], NVSS:[0.0015 0.0116 0.0217 0.0319], FIRST not available.
}}
\label{fig:c}
\end{figure}
{\bf RAD-Rabbit:} This unusual radio source, which resembles the face of a rabbit, was identified during the RAD@home Discovery Camp at the Institute of Physics (Bhubaneswar, Date: 23-29 June 2014). As shown in Figure.~\ref{fig:c} this ``Rabbit'' (RA: 12:19:34.95 Dec: -13:26:11.7, redshift $z=$0.06950) shows the host galaxy to be offset to the north from most of the radio emission seen in the TGSS data. This is neither consistent with an FR I nor an FR II structure and is likely a relic lobe (Figure.~\ref{fig:c} left panel). This source was further followed up with the GMRT (GOOD-RAC project) and found to show a diffuse/relic structure indeed. The east-west linear size is nearly 400 kpc with the radio core well-identified in the new GMRT data. The TGSS-NVSS spectral index ranges from -1.4 to -1.8, confirming its relic nature. The host galaxy also shows a merging companion in the Pan-STARRS images and that may have contributed to the highly disturbed circumgalactic environment distorting the radio lobe formation and/or evolution into a relic lobe.

{\bf RAD-giant DDRG:} This unusual radio source was identified during the Discovery Camp at UM-DAE Centre for Excellence in Basic Sciences, University of Mumbai (Mumbai, Date: 1-8 June 2014). This source showed a compact emission beside a large diffuse emission, linear in NVSS and curved in TGSS DR5 (but straight in ADR1). The compact radio source, suspected to be the core, did not show any optical counterpart in DSS optical data suggesting a high-z galaxy (Figure.~\ref{fig:c} right panel). If that is associated with the large diffuse source, it would be giant.
This was followed up through the GOOD-RAC project and found to be an episodic or a DDRG. The compact source remained compact but in the middle of the linear diffuse emission, a mini double lobe radio source was identified. The mini-double was well aligned with the larger radio source and on either side of the central optical galaxy (RA: 11:11:42.66, Dec: -13:24:11.43, Photo-z $=$ 0.113$\pm$0.018). Thus the relic lobe of this DDRG turned out to be a giant 800 kpc size.

%\subsection{RAD-Thumbs Up}

\begin{figure}
\centering
\includegraphics[width=0.5\columnwidth]{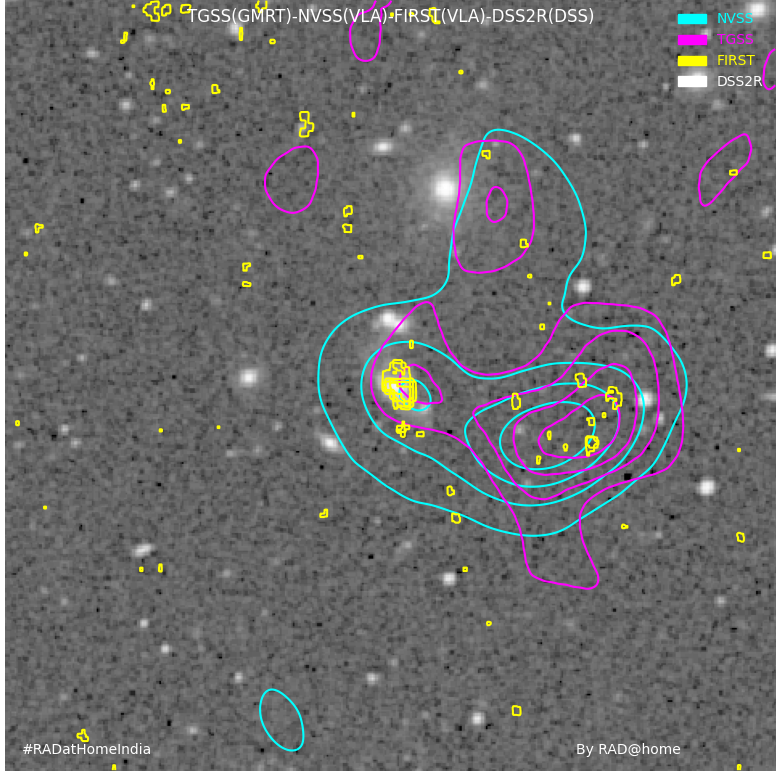}
\caption{\small{Left panel (RAD-Thumbs Up): Contour levels (Jy beam$^{-1}$) are TGSS:[0.015, 0.027, 0.038, 0.05 ], FIRST: [0.0005, 0.0007, 0.0009, 0.0011], NVSS:[0.0015, 0.004, 0.0065, 0.009].
}}
\label{fig:d}
\end{figure}

\begin{figure}
\centering
\includegraphics[width=\columnwidth]{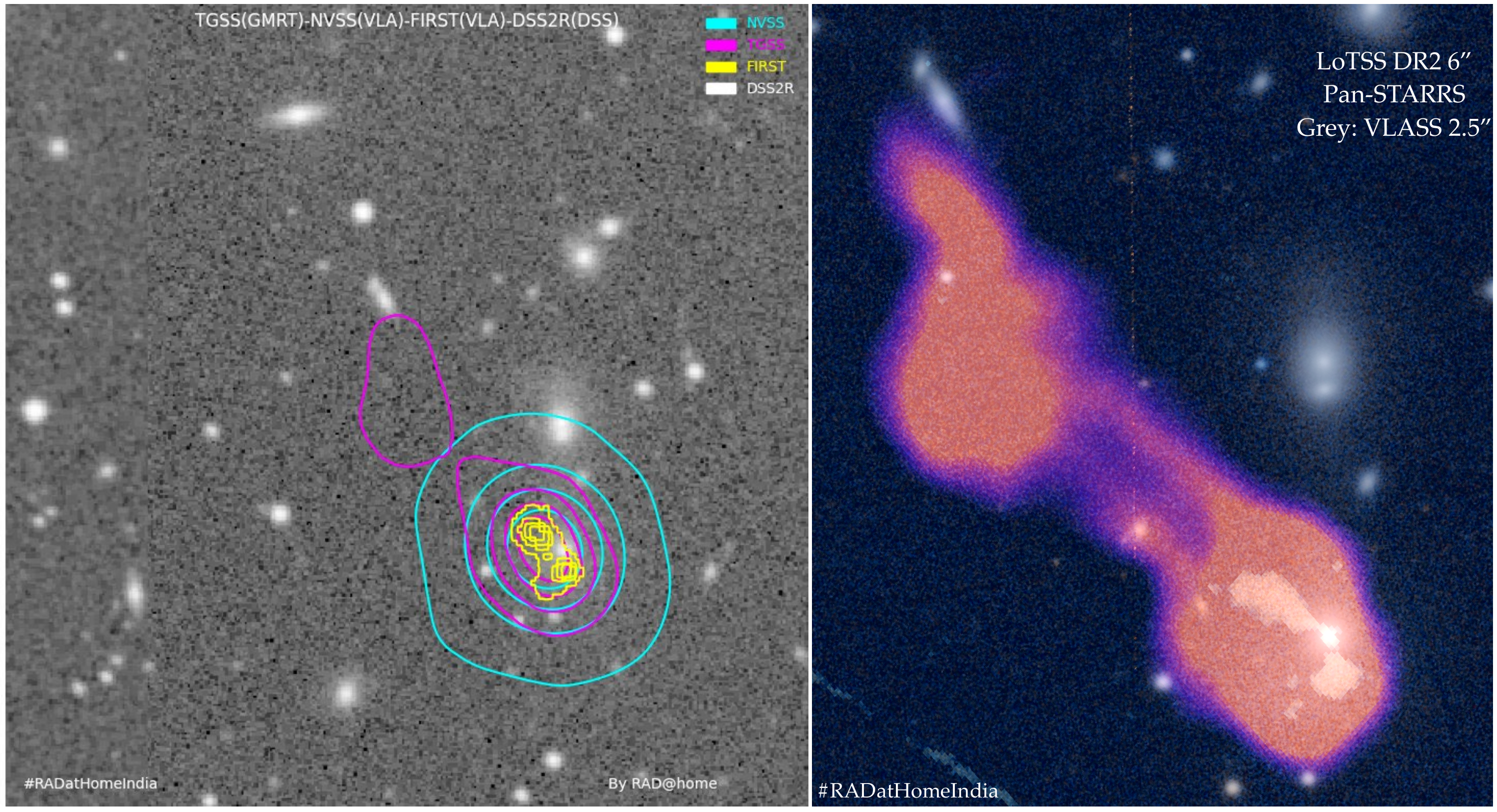}
\caption{\small{Left panel (RAD-Episodic WAT): Contour levels (Jy beam$^{-1}$) are TGSS:[0.015, 0.047, 0.079, 0.11], FIRST:[0.0005, 0.0018, 0.0032, 0.0045], NVSS:[0.0015, 0.0113, 0.0211, 0.0308]. Right panel: LoTSS DR2 144 MHz with 6\arcsec angular resolution in inferno colour, in grey VLASS 3 GHz with 2.5\arcsec angular resolution overlaid on Pan-STARRS image. This image has been made using the Aladin web tool. The VLASS image clearly resolves the inner structure of the radio galaxy and the LoTSS image shows the overall large extent of the source.
}}
\label{fig:d1}
\end{figure}

{\bf RAD-Thumbs Up:}This unique radio source was identified during the Discovery Camp at the Nehru Planetarium (Delhi; date: 22-26 December 2018). As can be seen in the RAD-composite contour image (Figure.~\ref{fig:d} ) where the NVSS (cyan) and TGSS (magenta) contours resemble the ``Thumbs up'' sign whose host galaxy was identified with a disturbed spiral galaxy (RA: 22:16:56.59 Dec: -04:24:33.4, $z=$0.095963 ). The host galaxy was confirmed with core and mini radio lobe detection in FIRST (yellow contours) confirming the nature of the diffuse structure to the west of the galaxy. Although the mini-lobe is seen to be comparable to stellar disk size (15 kpc) the size of the diffuse/relic-lobe emission is nearly 100 kpc, comparable to typical radio galaxies hosted in elliptical galaxies. However, this host galaxy is clearly a disturbed, young star-forming (UV-bright) spiral and even more strangely both the large-scale relic lobes are located on one side of the host in resemblance to wide-angle-tailed radio galaxies. Such large radio sources associated with spirals and leaving both the relic lobes on one side is extremely rare, if not unique. Although the linear size is small, an analogy can be drawn to the Virgo cluster spiral NGC4438 where a mini-double along with diffuse relic emission on one side can be seen due to ram pressure stripping \cite{Hota2007}.

%\subsection{RAD-Episodic WAT} 

{\bf RAD-Episodic WAT:} This interesting source was identified (Date: 28th October 2017) during weekend online e-classes discussion sessions conducted by the collaboratory. Comparing all three radio data and optical data (as shown in the composite contour image Figure.~\ref{fig:d1} left panel) it can be seen that the compact radio lobes ($\sim$\,50 kpc size) are located on either side of the host galaxy (RA: 15:33:20.24 Dec: +31:08:50.2, $z=$0.066694 ) which also coincide with the NVSS and TGSS peaks. However, a faint emission tail, 100-200 kpc size, is seen in the TGSS to the north-east of the galaxy which is not detected in NVSS/FIRST suggesting the diffuse and steep spectral nature. Recent LoTSS data [rms noise 80 $\mu$Jy per beam (6\arcsec)] confirms this diffuse tail with a great significance level (Figure.~\ref{fig:d1} right panel). The natural explanation would be an episodic nature along with ram pressure stripping by the group/cluster medium which is strong enough to push back the relic but not enough to disrupt mini radio lobe formation. Such an environment gives rise to wide-angle-tailed (WAT) radio galaxies. Such episodic feedback along with ram pressure stripping during infall of galaxies into clusters may be depriving the galaxies of fuel for future star formation. 
  
%\subsection{RAD-one sided relic lobe}

\begin{figure}
\centering
\includegraphics[width=\columnwidth]{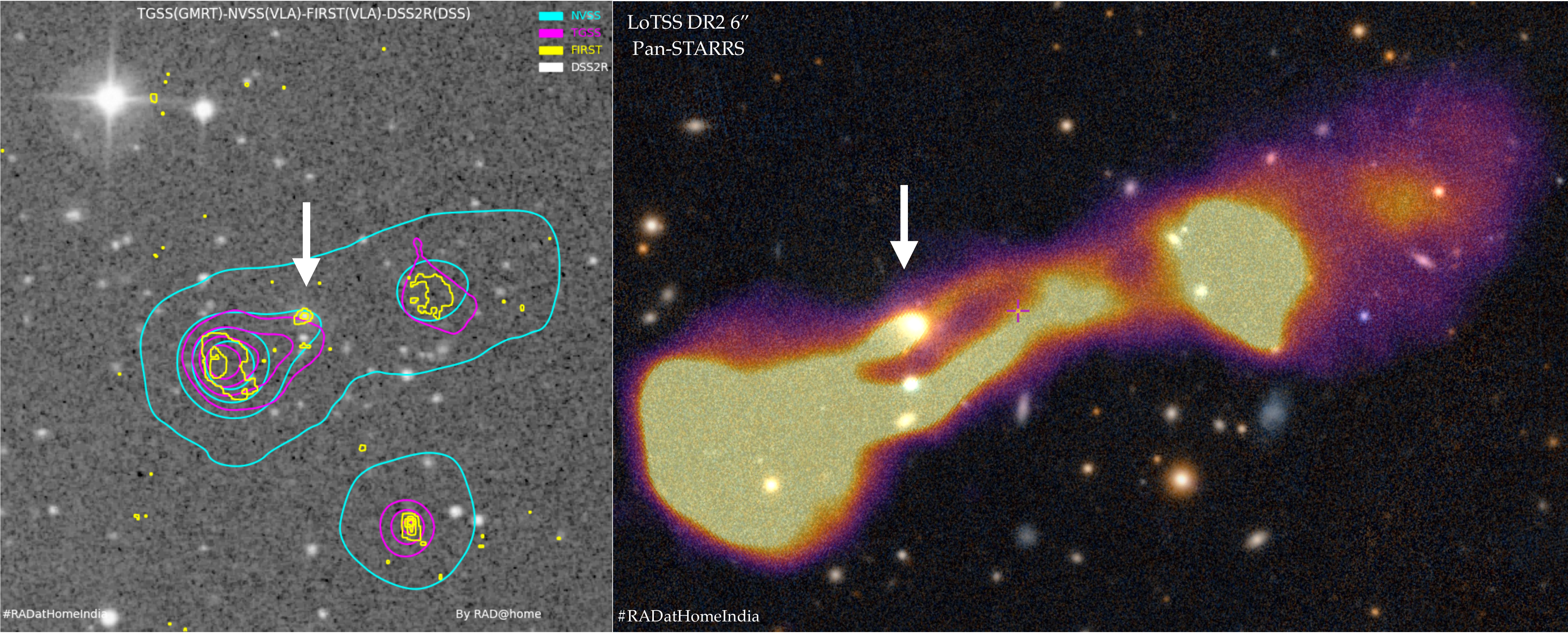}
\caption{\small{ Shows RAD-Collimated Synchrotron Thread source. Left Panel: Contour levels (Jy beam$^{-1}$) are TGSS:[0.015, 0.036, 0.057, 0.078], FIRST: [0.0005, 0.0028, 0.0051, 0.0074], NVSS:[0.0015, 0.0181, 0.0346, 0.0512]. Right panel: LoTSS DR2 144 MHz with 6\arcsec angular resolution in inferno colour overlaid or multi-colour Pan-STARRS image. The host galaxy has been marked with an arrow sign on both panels. This image has been made using the Aladin web tool.}}
\label{fig:e1}
\end{figure}

%\subsection{RAD-Collimated Synchrotron Thread}

{\bf RAD-Collimated Synchrotron Thread:} Diffuse NVSS emission extending beyond compact emission structures seen in FIRST is a suggestion for relic lobes if the radio source is similar to FR II radio galaxy. Such a case was found during e-class (Date: 19-21 October 2023). The radio core (RA: 08:31:09.55 Dec:+41:47:38.41), eastern lobe and western lobe were identified in FIRST data but only on the western side did the NVSS emission extend farther west (left panel of Figure.~\ref{fig:e1}). Similar to previously described cases, if the host galaxy is in motion towards the east, it is possible to explain the structure with two minor episodes of activity where the old and young lobes overlap on the east but are seen separated on the west. The LoTSS DR2 data  (right panel of Figure.~\ref{fig:e1}) additionally reveals an entirely new linear radio structure connecting the eastern and western lobe, seen south to the radio core. This makes it a rare case of so-called ``collimated synchrotron threads'' seen remarkably with the MeerKAT telescope in the galaxy ESO 137-006 \cite{Ramatsoku2020}. We suggest satellite galaxies moving around the host may be creating magnetic channels and these are being energised by the plasma from the radio lobes. A detailed study of the target is in progress.
%\subsection{RAD-Giant-radio spiral}

\begin{figure}
\centering
\includegraphics[width=0.4\columnwidth]{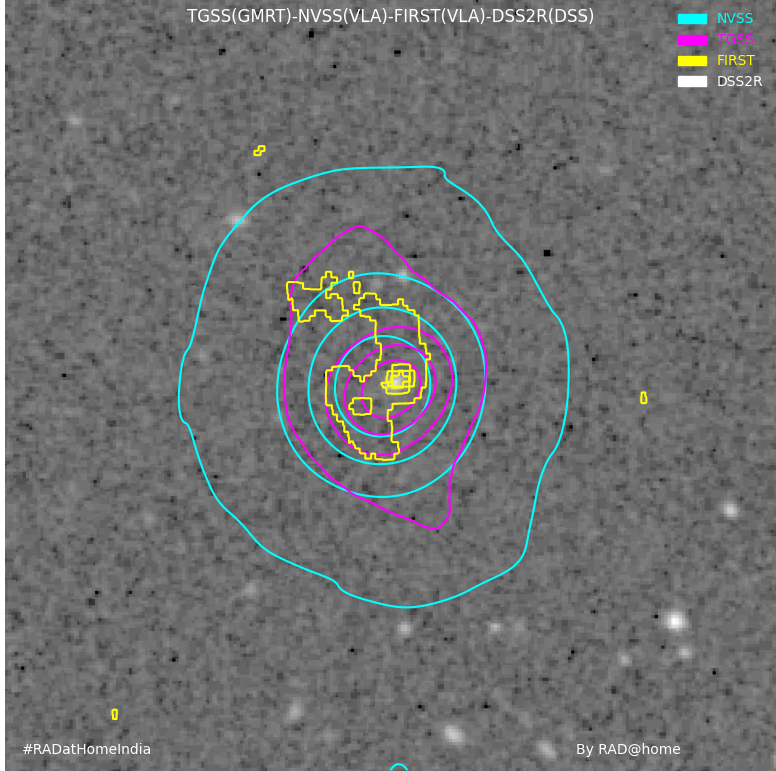}
\includegraphics[width=0.325\columnwidth]{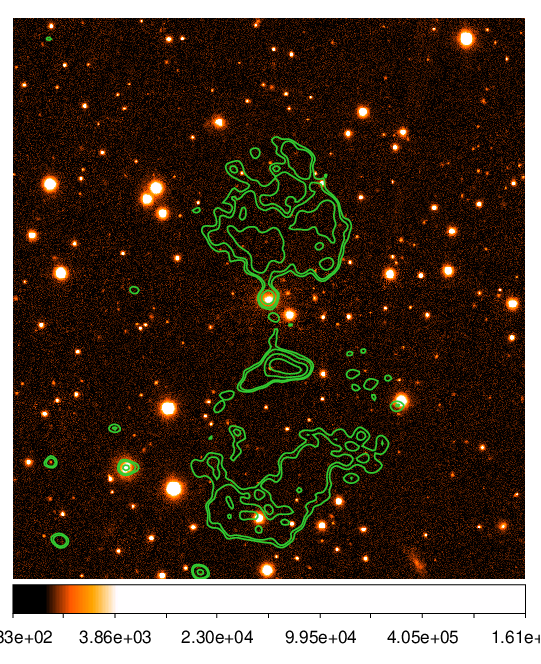}
\caption{\small{Left panel (RAD-Giant-radio spiral):Contour levels (Jy beam$^{-1}$) are TGSS:[0.015, 0.1, 0.186, 0.271], FIRST:[0.0005, 0.0093, 0.0181, 0.0269], NVSS:[0.0015, 0.0291, 0.0567, 0.0843]. Right Panel (RAD-double-bubble): LoTSS contour levels in square root scale with values (starting with 3 x rms) are 0.00024, 0.00031875, 0.000555, 0.00094875, 0.0015 Jy beam$^{-1}$. The background false-colour image is from Pan-STARRS in i-band.
}}
\label{fig:f}
\end{figure}

{\bf RAD-Giant-radio spiral:} It is normal to resolve a point source of NVSS to a double-lobe radio source in FIRST data due to better angular resolution (45\arcsec  ~v/s  ~5\arcsec ). However, in the following unusual case, spotted on 18th October 2023 the higher resolution data revealed a spiral-galaxy-like structure (Figure.~\ref{fig:f} left panel) which is neither related to the star-forming arms of the host galaxy (SDSS J133116.48+441851.4, Photometric $z=$0.250$\pm$0.0515 ) nor the radio lobes are sub-galactic scale where the rotating ISM-could bend the lobes (e.g. NGC3801; \cite{Hota2012}). In Pan-STARRS images the host galaxy is clearly seen as an edge-on disk/lenticular galaxy, roughly aligned east-west. The source structure is confirmed from the LoTSS images where the spiral structure extends nearly 450 kpc, larger than typical FR I and FR II radio galaxies. 
% In analogy with NGC3801 and CenA, it can be speculated that the host galaxy is a merger-remnant and/or a kinematically decoupled core galaxy where the nuclear gas/dust disk is orthogonal to the large-scale gas disc. 
Analogous to NGC 3801 and CenA, it is possible that the host galaxy is either a merger remnant or a kinematically decoupled core galaxy, with the nuclear gas/dust disk orthogonal to the large-scale gas disk.
In future, sensitive SKAO images are likely to reveal more such radio-spirals and help understand the interaction between radio jets and circum-galactic medium which is not possible with other tracers of atomic, molecular and ionised gas.

{\bf RAD-double-bubble:} Since TGSS (25\arcsec) has a higher resolution than NVSS (45\arcsec) some relic lobes can be missed in interferometric TGSS images. So citizen scientists started checking the availability of some of their potential relic lobe cases in recent LoTSS data. One such case resembles two bubble-like lobes, found during e-class (Date: 20 September 2023), which is presented here. Identifying the host galaxy for such sources also becomes challenging, as radio cores may not be visible. However, due to the availability of sensitive surveys like Pan-STARRS host galaxies can now be easily identified from colour images. The RGB-maker tool is yet to be updated to add such higher sensitivity images and hence, the SAO ds9 image analysis tool as used by the citizen scientists along with NED and Aladin. The relic lobe presented here neither corresponds to an edge-darkened FR I nor an edge-brightened FR II, rather diffuse bubbles (Figure.~\ref{fig:f} right panel). The host galaxy is associated with a point source (RA:02:26:31.68, Dec:+42:09:03.28, $z=$0.144). Hence, the total north-south extent of the relic lobes, 250\arcsec, corresponds to 635 kpc, a giant radio galaxy. In cases when the core is undetected or lobe orientation is well-aligned, it may also become classified as the puzzling class of objects known as Odd Radio Circles \cite{Norris2021}.

\begin{figure}
\centering
\includegraphics[width=\columnwidth]{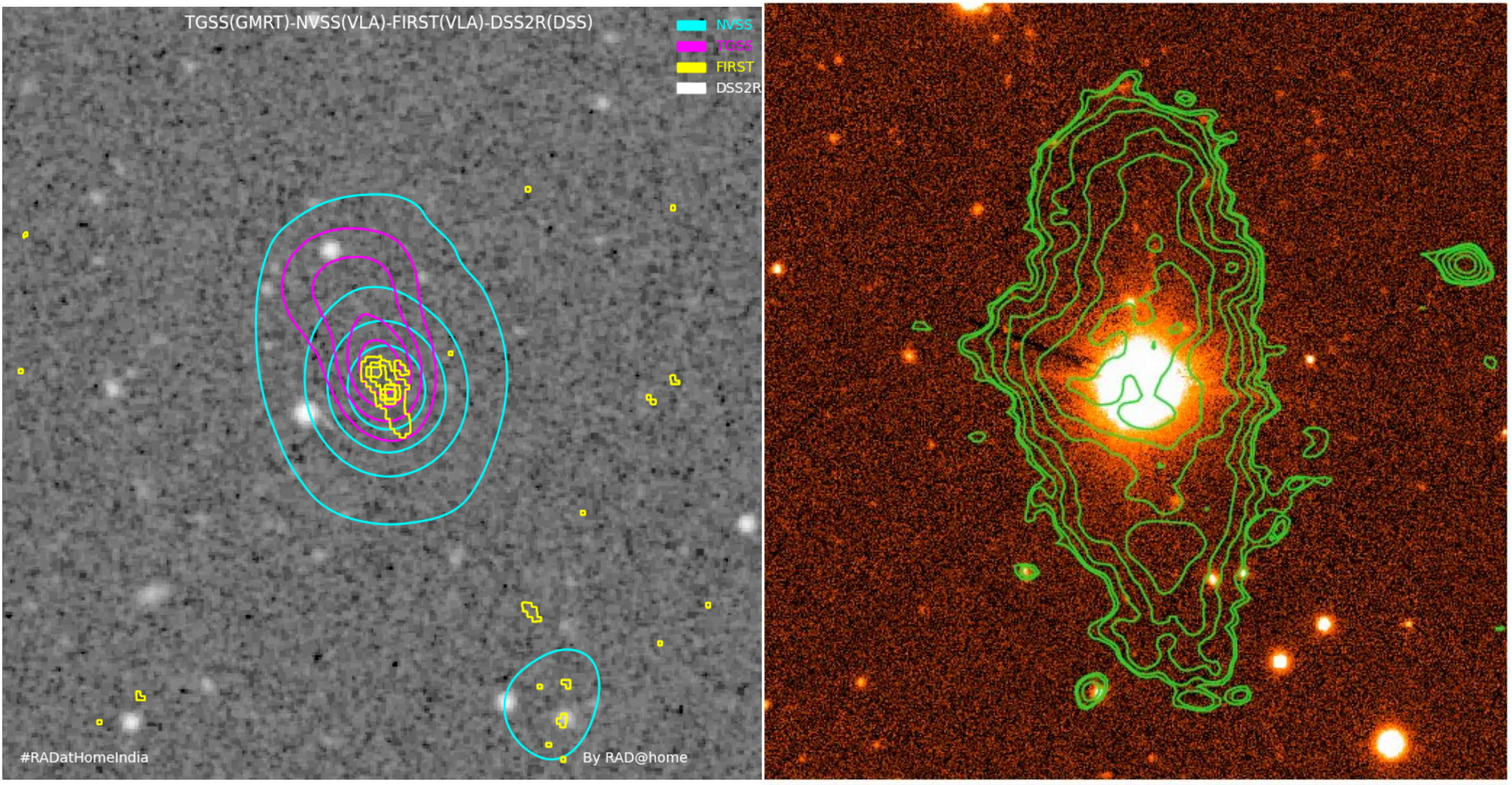}
\caption{\small{Left panel (RAD-one sided relic lobe): Contour levels (Jy beam$^{-1}$) are TGSS:[0.015, 0.029, 0.043, 0.057], FIRST:[0.0005, 0.0025, 0.0044, 0.0064], NVSS:[0.0015, 0.0078, 0.0141, 0.0203]. Right panel: LoTSS contour levels in square root scale with values (starting with 3 $\times$ rms) are 0.00024, 0.000275918, 0.000383673, 0.000563265, 0.000814694, 0.00113796, 0.00153306, 0.002 Jy beam$^{-1}$. The background false-colour image is from Pan-STARRS in i-band.
}}
\label{fig:g}
\end{figure}

{\bf RAD-one-sided relic lobe:} Yet another similar case was found during an online e-class (Date: 23-28 October 2023). The point source of the NVSS image was resolved to a triple radio source in the FIRST data with a radio core coinciding with the optical host galaxy (RA: 12:19: 11.51 Dec: +61:08:16.7, $z=$0.552). However, only TGSS shows a clear northern extension suggesting a relic lobe (Figure.~\ref{fig:g} left panel). Recent LoTSS data resolves the TGSS tail to be a clear relic lobe of linear size near 425 kpc. This relic blob is bigger than the compact lobes of 230 kpc. A resolved imaging study of the relic lobe in any frequency other than around 150 MHz is needed to better understand it. The overall structure suggests a similarity with the previous episodic WAT. Thus it is likely a cluster-infalling galaxy, and episodes of feedback during infall may play an important role in the transformation of galaxies.

%\subsection{RAD-radio-fog-beyond-star}
{\bf RAD-radio-fog-beyond-star:} Identifying the host galaxy for any diffuse radio sources has always been challenging. The host galaxy may have moved away after active jet creation. A confusing case, found during e-class (Date: 13-15 October 2022), presented here has a bright point source in the optical data (BD+39 2856, High Proper Motion Star). Furthermore, unlike the above case where the relic lobes are well separated in space, suggesting the host galaxy may be in the geometrical mid-point where the midpoint itself has the star. Bright stars are poor emitters of low-frequency radio emission. With careful analysis of the radio structure, although it has a north-south elongation the middle region has a significant east-west extension, we can see only two sources as possible host galaxies (Figure.~\ref{fig:g} right panel). The source to the north (RA:15:16:19.91, Dec:+38:34:52.88, photometric $z=$0.192$\pm$0.064) and a fainter one on the north-west (RA:15:16:18.73, Dec:+38:34:46.52, photometric $z=$0.281$\pm$ 0.286) are two possible host galaxies. Thus the radio source could be nearly 300 kpc long which is typical of FR II radio galaxies. Our understanding of radio galaxies is prejudiced by bright nearby sources hosted in elliptical galaxies. Such radio sources, not fitting to any class, may reveal new modes (diffuse bubbles) of radio emission in the near future.

\section{Discussion and Summary}
We have briefly presented 11 interesting objects from our decade-old citizen science research programme named RAD@home. The participants were initially motivated to discover relic radio lobes as ``faint-and-fuzzy'' radio blobs in the TGSS 150 MHz radio survey images of the sky obtained using GMRT. In the process, the objects discovered were so remarkable that radio structures far beyond the typical features of jetted Seyferts and classical radio galaxies were identified. Burl-like features have been found on jets that may or may not be due to jet-galaxy interactions. Collimated Synchrotron Threads (CST) have been found that may be re-energised magnetic channels of satellite galaxies around the host radio galaxies. Radio spirals have been found which are larger than the typical FR II radio galaxies and they may have originated in kinematically decoupled core galaxies formed after a merger. Tailed infalling radio galaxies have been observed which show multiple episodes of activity. Some relic radio lobes form in the shape of bubbles, which may explain odd radio circles and some relic lobes can be purely diffuse almost without any structure.

Spiral galaxies exhibit a variety of continuum radio properties, ranging from diffuse radio halos without active galactic nuclei (AGN), such as M82 and NGC 253, to galaxy-scale radio bubbles associated with Seyfert nuclei (e.g., Circinus galaxy, NGC3079, \& NGC6764; \cite{Hota2006} and references therein). Additionally, some spiral galaxies exhibit relic or FR II radio lobes extending up to Mpc-scales which used to be associated only with giant elliptical galaxies, prior to the discovery of `Speca' in 2011 \cite{Hota2011}. Close to 3 dozen Speca-like galaxies have been discovered since then. This reflects a spectrum of low to high-power AGNs with varying manifestations of radio jets. Some underlying features or intrinsic properties of these galaxies may remain undetected or unobserved due to the limited sensitivity of current observational instruments. However, this limitation will be addressed with the advent of the Square Kilometre Array Observatory (SKAO), which will offer unprecedented sensitivity and superior observational capabilities. This advancement may also uncover details of radio-mode feedback acting on neutral atomic hydrogen gas (H{\sc i}) located far beyond the stellar light distribution. This will provide new insights into previously unexplored modes of AGN feedback affecting the circumgalactic medium (CGM) of a galaxy. While Artificial Intelligence (AI) and Machine Learning (ML) may lead source classification in large datasets, citizen science research—particularly the Collaboratory model like RAD@home—can provide novel directions and handle complex cases more effectively. Furthermore, it prepares the next generation of astronomers, discovers new or peculiar types of sources (which serve as starting points for training AI/ML models), and contributes towards sustainable development goals (\url{https://sdgs.un.org/goals}).

\begin{acknowledgement}
AH thanks the ISRA2023 conference organisers for the wonderful hospitality and environment conducive to academic discussions. AH acknowledges the University Grants Commission (UGC, Ministry of Education, Government of India) for monthly salary for the last 11 years. The long list of national and international organisations which have helped the establishment and growth of the first Indian astronomical citizen science research platform, RAD@home, are acknowledged in detail at \url{https://radathomeindia.org/brochure}.  
\end{acknowledgement}

%\bibliographystyle{spphys}
%\bibliography{bondirefs}

\end{document}